\documentclass{article}
\usepackage{emulateapj,pstricks,apjfonts}

\def\asca	{{\em ASCA}\/}
\def\ginga	{{\em Ginga}\/}
\def\rosat	{{\em ROSAT}\/}
\def\chandra 	{{\em Chandra}\/}
\def\xmm 	{{\em XMM}\/}
\def\bepposax 	{{\em BeppoSAX}\/}
\def\am		{$^\prime$}
\def\as		{$^{\prime\prime}$}
\def\ergscm	{~erg$\;$s$^{-1}\,$cm$^{-2}$}
\def\kmsmpc	{~km$\;$s$^{-1}\,$Mpc$^{-1}$}
\def\cmsq	{~cm$^{-2}$}
\def\lax{\lesssim}
\def\bi{\bfseries\itshape}

\begin{document}

\submitted{ApJ in press; astro-ph/0105093 v2}

\lefthead{MERGER SHOCKS IN A665 AND A2163}
\righthead{MARKEVITCH ET AL.}

\title{MERGER SHOCKS IN GALAXY CLUSTERS A665 AND A2163 AND THEIR RELATION TO
RADIO HALOS}

\author{M. Markevitch and A. Vikhlinin\altaffilmark{1}}

\affil{Harvard-Smithsonian Center for Astrophysics, 60 Garden St.,
Cambridge, MA 02138; maxim, alexey @head-cfa.harvard.edu}

\altaffiltext{1}{Also Space Research Institute, Russian Academy of Sciences}

\begin{abstract}

We present \chandra\ gas temperature maps for two hot, intermediate-redshift
clusters A665 and A2163. Both show strong temperature variations in their
central $r=0.5\,h^{-1}$ Mpc regions, naturally interpreted as product of the
subcluster mergers. The A665 map reveals a shock in front of the cool core,
while the temperature structure of A2163 is more complicated.  On a larger
linear scale, our data on A2163 indicate a radial temperature decline in
agreement with earlier \asca\ results, although the uncertainties are large.
Both these clusters exhibit previously known synchrotron radio halos.
Comparison of the radio images and the gas temperature maps indicates that
radio emission predominantly comes from the hot gas regions, providing a
strong argument in favor of the hypothesis that relativistic electrons are
accelerated in merger shocks.

\end{abstract}


\section{INTRODUCTION}

The intracluster medium (ICM) is heated to the observed high temperatures by
gas-dynamic shocks when the cluster is formed through infall and merging of
smaller subunits. In clusters currently undergoing mergers, such shocks may
be identified in the gas temperature and density maps, if the geometry of
the merger is favorable. The ICM density and temperature can be derived from
X-ray data. Due to the difficulty of the spatially-resolved X-ray
spectroscopy, so far only a few nearby clusters have been studied in
sufficient detail, but a number of them indeed exhibit the characteristic
irregular temperature structure (\rosat\ and \asca\ results, e.g., Henry \&
Briel 1995; Markevitch, Sarazin \& Vikhlinin 1999 and references therein;
\chandra\ and \xmm\ results, e.g., Markevitch et al.\ 2000; Vikhlinin,
Markevitch, \& Murray 2001; Neumann et al.\ 2001).

At radio frequencies, some clusters exhibit large, centrally located, low
surface brightness halos with relatively steep spectra (for historic
references see, e.g., Sarazin 1988; recent works, e.g., Giovannini, Tordi,
\& Feretti 1999 and references therein). Such halos are relatively rare.
They are generated by a population of ultra-relativistic electrons
(coexisting with thermal electrons of the ICM) that emit synchrotron
radiation in the cluster magnetic field. Such electrons should be relatively
short-lived ($\sim 10^8$ yr) due to inverse Compton and synchrotron energy
losses (see Sarazin 1999 for detailed modeling), and yet they must have time
to spread over the cluster volume or they should be accelerated in situ. The
source of such electrons that can support a halo for a sufficient time is
unclear. Several possibilities were proposed, including radio galaxies
(Jaffe 1977), interaction of cosmic ray protons with the ICM protons (Dolag
\& En{\ss}lin 2000 and references therein), and the turbulence generated by
merger shocks (Harris, Kapahi, \& Ekers 1980; Tribble 1993). Indeed, merger
shocks dissipate vast amounts of kinetic energy that, besides heating the
intracluster gas, may power the amplification of magnetic fields and
acceleration (or re-acceleration) of relativistic particles. A correlation
between the existence of a radio halo and the irregular cluster shape (that
usually indicates a merger) has been noticed (e.g., Feretti 2000 and
references therein). Further, Buote (2001) showed that for clusters with
known radio halos, the radio power correlates with the degree of the X-ray
morphological disturbance. Such correlations favor the merger origin of
halos. On the other hand, there are a few highly disturbed clusters without
halos (Buote 2001) and, as pointed out by Liang et al.\ (2000), halos in
relaxed clusters may have escaped detection, because such clusters usually
have strong central radio sources making observation of a low surface
brightness halo technically difficult. A correlation between the radio halo
power and the cluster gas temperature was reported (Liang et al.\ 2000 and
references therein). It was also noticed for several well-resolved halos
that the radio brightness follows the X-ray brightness on large scales
(e.g., Deiss et al.\ 1997; Govoni et al.\ 2001).  Obviously, the best check
of the significance of merger shocks for the radio halo formation would be
to find clusters with shocks and halos and see if there is any relation
between the shock location and the radio emission.

In this paper, we analyze short \chandra\ ACIS observations of two hot
clusters A665 ($z=0.182$) and A2163 ($z=0.201$). Their X-ray morphology was
studied with earlier telescopes, most recently with \rosat\ (e.g., Buote
\& Tsai 1996; G\'omez, Hughes, \& Birkinshaw 2000; Elbaz, Arnaud, \&
B\"ohringer 1995), and the crude gas temperature maps were derived with
\asca\ (Markevitch et al.\ 1994; Markevitch 1996). Those data suggested that
both clusters are undergoing mergers. Using the new \chandra\ data, we
derive the gas temperature maps at a much finer linear scale and locate the
sites of shock heating of the ICM. Both these clusters have giant radio
halos (discovered in A665 by Moffet \& Birkinshaw 1989 and in A2163 by
Herbig \& Birkinshaw 1994). For the first time for any radio halo clusters,
we compare the radio images and the gas temperature maps. We use
$H_0=100\,h$\kmsmpc\ and $\Omega_0=0.3$; confidence intervals are
one-parameter 90\%.

%%%%%%%%%%%%%%%%%%%%%%%%%%%%%%%%%%%%%%%%%%%%%%%%%%%%%%%%%%%%%%%%%%%%%%%%%%
\begin{figure*}[tb]
\pspicture(0,10.2)(18.5,18.8)
%\psgrid(0,10)(18.5,20)

\rput[tl]{0}(0.8,20){\epsfxsize=8cm
\epsffile{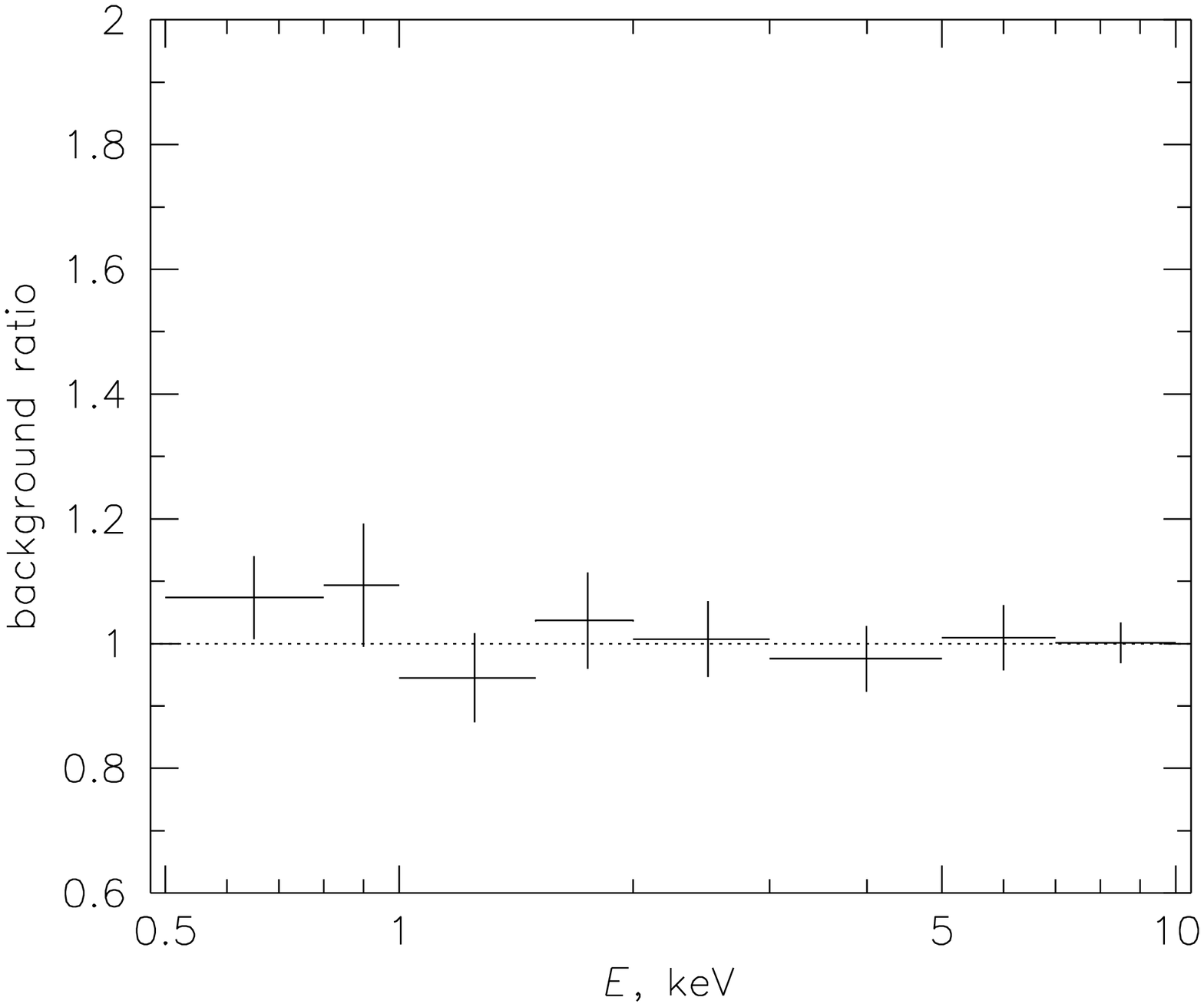}}

\rput[tl]{0}(9.8,20){\epsfxsize=8cm
\epsffile{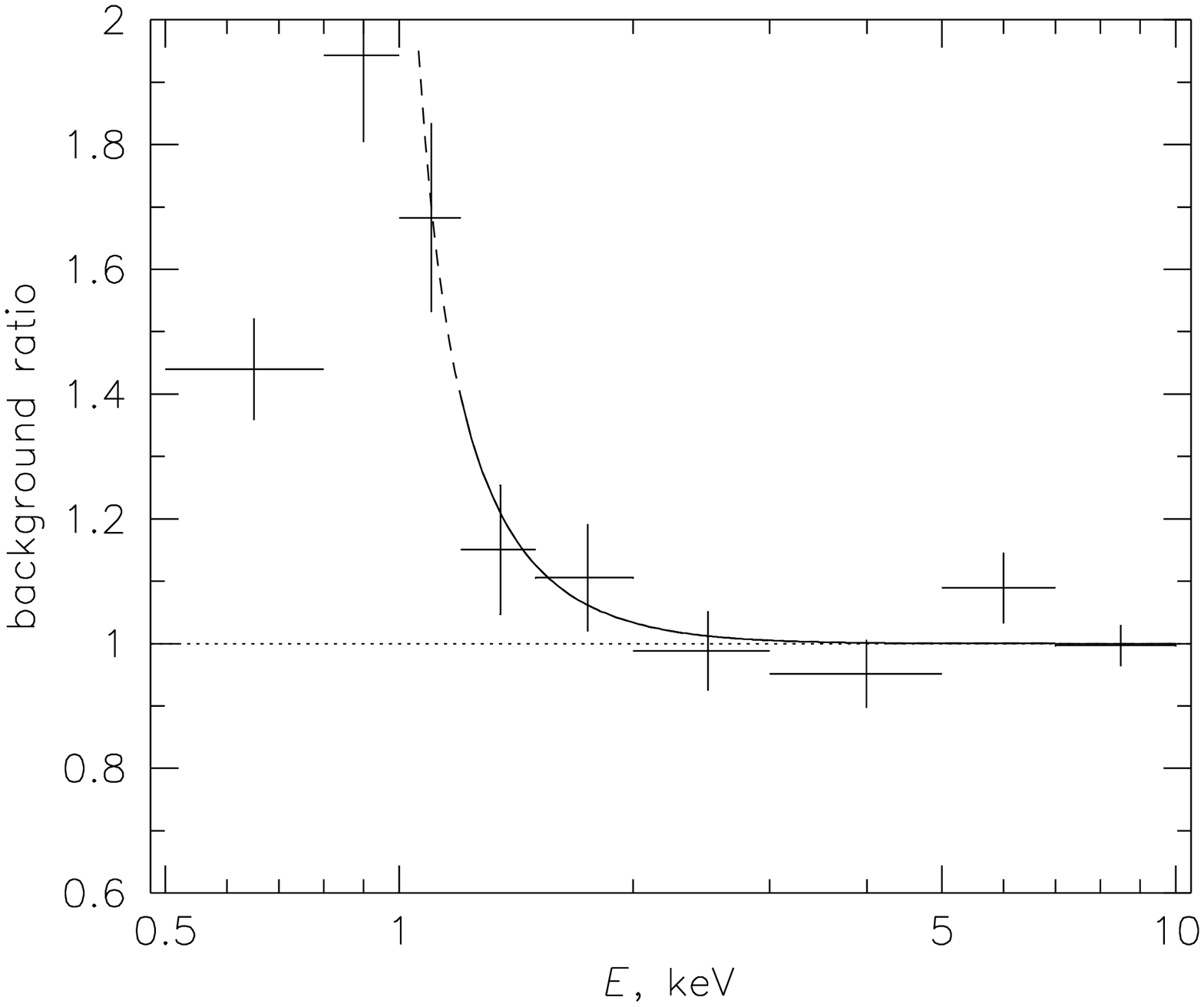}}

\rput[bl]{0}(7.8,17.9){\large\bi a}
\rput[bl]{0}(16.8,17.9){\large\bi b}

\rput[tl]{0}(-0.1,11.7){
\begin{minipage}{18.5cm}
\small\parindent=3.5mm
{\sc Fig.}~\ref{bgratios}.---Ratio of the observed background rate in chip
S2 (that has negligible cluster emission) to the expected rate from the
blank-field data corrected for a long-term trend, ({\em a}\/) for A665 and
({\em b}\/) for A2163. Vertical errors are $1\sigma$.  For A2163, an
approximate analytic description of the ratio is shown (the line is solid in
the $E>1.2$ keV interval that is used for spectral fits). This soft excess
can also be described by a thermal emission arising from a warm foreground
gas (our preferred method, see text).
\par
\end{minipage}
}
\endpspicture
\refstepcounter{figure}
\label{bgratios}
\end{figure*}
%%%%%%%%%%%%%%%%%%%%%%%%%%%%%%%%%%%%%%%%%%%%%%%%%%%%%%%%%%%%%%%%%%%%%%%%%%

\section{DATA ANALYSIS}
\label{sec:data}

A665 was observed on 1999 December 29 with ACIS-I%
\footnote{\chandra\ Observatory Guide
http://asc.harvard.edu/udocs/docs/, ``Observatory Guide,'' ``ACIS''}
using chips I0--I3 and S2.  The useful exposure was 8550 s; there were no
background flares (Markevitch 2001) during the observation. A2163 was
observed on 2000 July 29 using the same ACIS configuration. The useful
exposure was 9560 s and there were no background flares. Events with bad
grades and those originating from bad pixels and columns were screened out
in a standard manner.

%%%%%%%%%%%%%%%%%%%%%%%%%%%%%%%%%%%%%%%%%%%%%%%%%%%%%%%%%%%%%%%%%%%%%%%%%%
\begin{figure*}[tb]
\pspicture(0,10.5)(18.5,19.4)
%\psgrid(0,10)(18.5,20)

\rput[tl]{0}(0.8,20){\epsfxsize=8.5cm
\epsffile{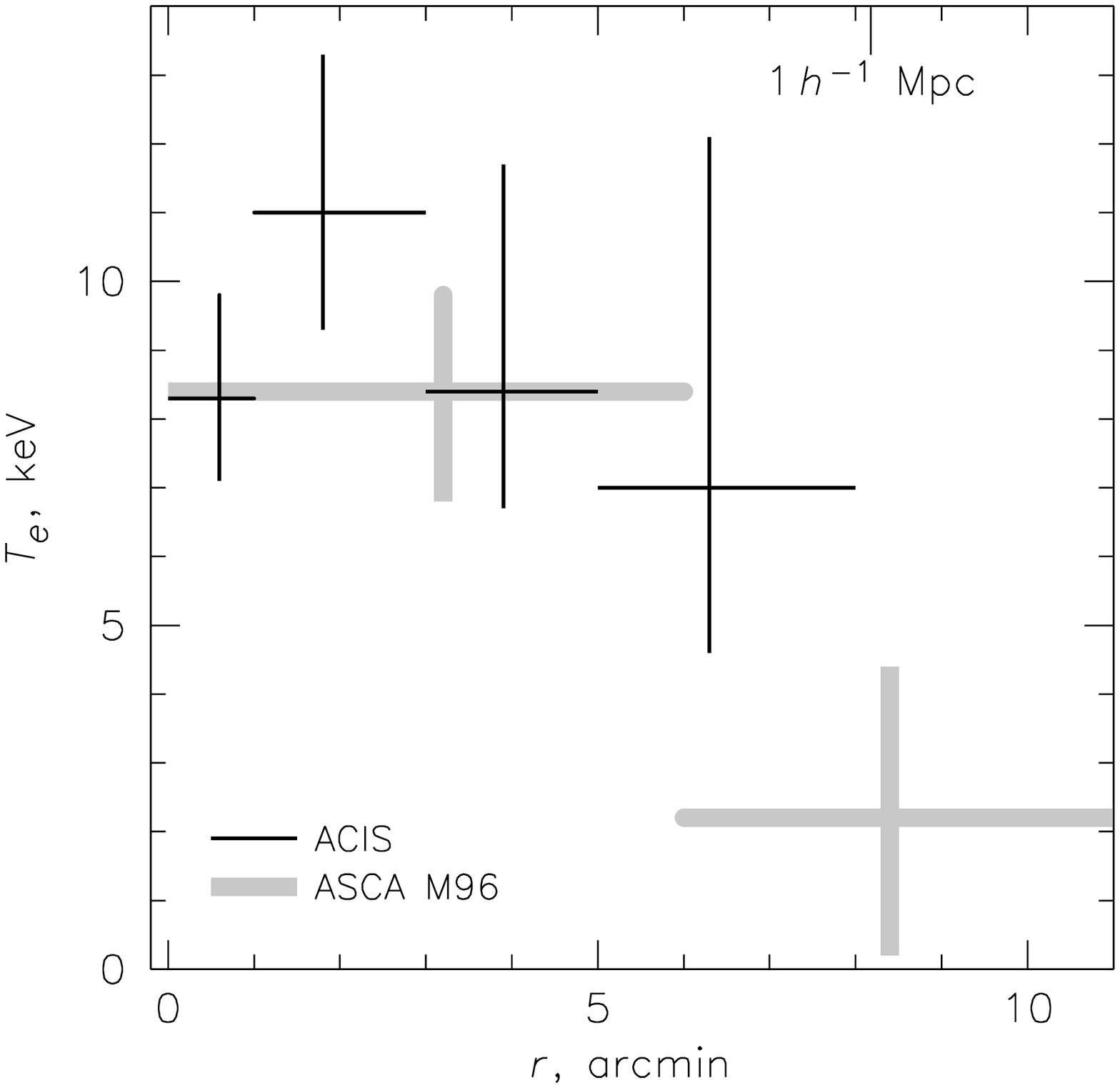}}

\rput[tl]{0}(9.3,20){\epsfxsize=8.5cm
\epsffile{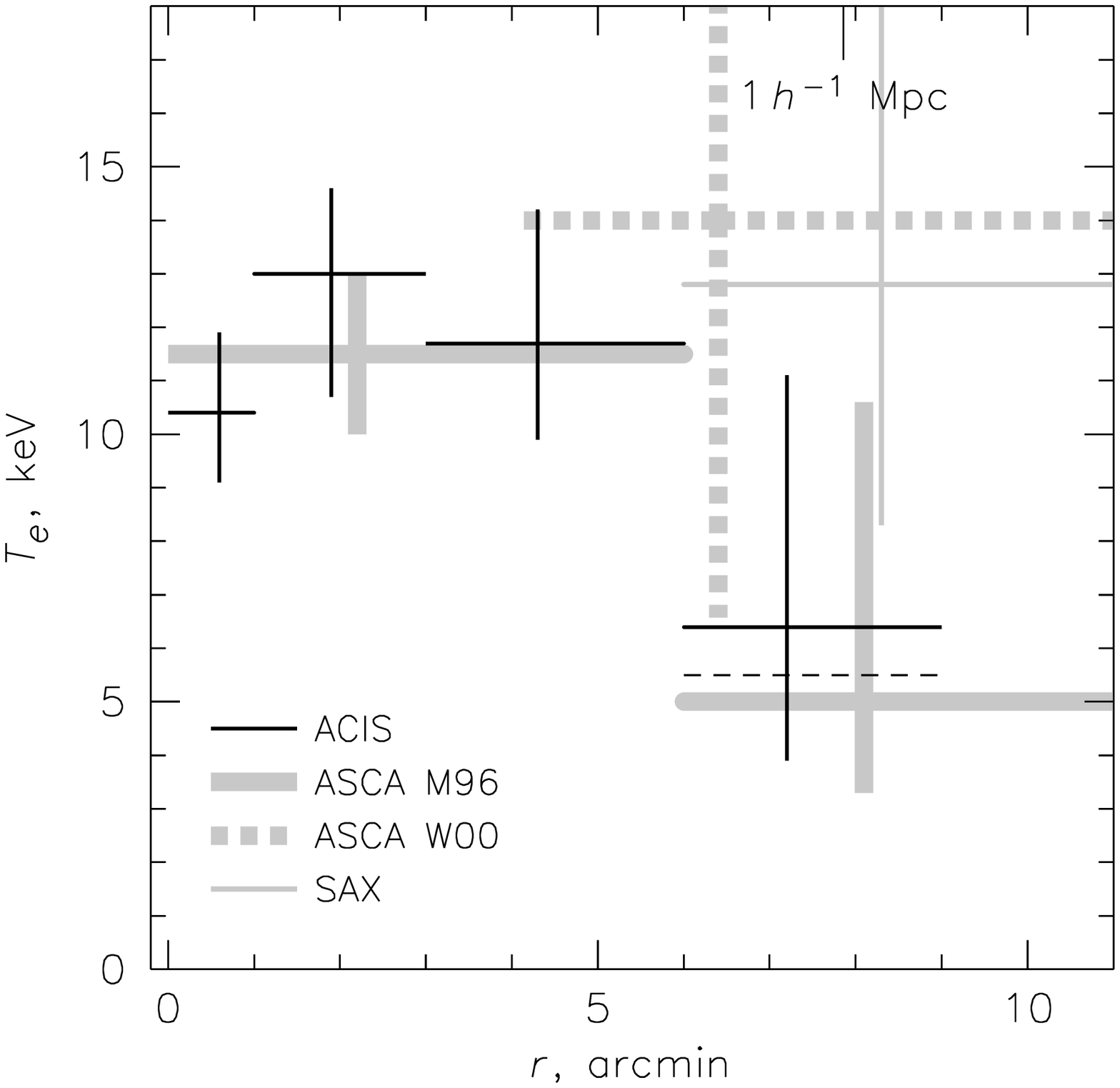}}

\rput[bl]{0}(2.5,18.6){\large\bi a}
\rput[bl]{0}(11.1,18.6){\large\bi b}

\rput[tl]{0}(-0.1,11.6){
\begin{minipage}{18.5cm}
\small\parindent=3.5mm
{\sc Fig.}~\ref{tprofs}.---Radial projected temperature profiles, ({\em
a}\/) for A665 and ({\em b}\/) for A2163. Black lines show results obtained
here, while gray lines show earlier \asca\ results from Markevitch (1996)
and White (2000) and \bepposax\ results from Irwin \& Bregman (2000).
Uncertainties are 90\% and crosses are centered at the emission-weighted
radii. Black dashed line in panel ({\em b}\/) shows the best-fit temperature
value with the soft background excess ignored (see text).
\par
\end{minipage}
}
\endpspicture
\refstepcounter{figure}
\label{tprofs}
\end{figure*}
%%%%%%%%%%%%%%%%%%%%%%%%%%%%%%%%%%%%%%%%%%%%%%%%%%%%%%%%%%%%%%%%%%%%%%%%%%

\subsection{Background subtraction}
\label{sec:bg}

Background subtraction is critical for deriving the temperatures in the
low-brightness regions at high cluster radii, which is one of the goals of
this work. The main component of the background is modeled using the
composite blank-field datasets%
\footnote{Described at http://hea-www.harvard.edu/$\sim$maxim/axaf/acisbg}
for the corresponding focal plane temperature, normalizing them by the ratio
of the exposures and a small energy-independent correction factor that
reflects the long-term decrease of the background (0.95 for A665 and 0.93
for A2163, determined by the observation dates). The background datasets
include mostly the fields with low Galactic $N_H$, appropriate for most
clusters including A665. However, A2163 has a relatively high $N_H$ (there
is also an unusual difference between the Galactic H~{\small I} measurements
and the best-fit \rosat\ PSPC value, see Elbaz et al.\ 1995), thus we should
verify the applicability of the standard background. In both observations,
we can compare the background model to the data from chip S2 that covers a
region sufficiently far from the cluster centers to ignore the cluster
contribution. Figure~\ref{bgratios} shows the ratios of the spectra from
chip S2 to the spectra from the respective blank field datasets with proper
normalizations. For A665, the model predicts the observed S2 background
perfectly.

The A2163 observation, however, shows a clear soft excess, already noticed
by Pratt et al.\ (2001) in the \xmm\ observation of this cluster.  This
excess may be due to the local enhancement of the diffuse Galactic
background, or a temporary increase of the soft particle background at the
time of the observation, or a combination thereof. An excess is also seen in
the \rosat\ PSPC observation of this field, thus, given all the data, a
genuine enhancement in the sky is more likely. Examination of the \rosat\
PSPC image in the relevant 0.7--1.3 keV band shows no spatial variations of
the soft background in the vicinity of A2163 to within 5\%. Therefore, we
can model this excess by fitting its spectrum from chip S2 and then adding
the best-fit model, scaled by the region area, to the model of the cluster
emission (thus a small difference in vignetting between the cluster position
and the field covered by S2 will be taken into account properly).  After the
subtraction of the nominal blank-field background, the S2 excess spectrum in
the 0.5--3 keV band is described quite well by a simple absorbed thermal
model (MEKAL, Kaastra 1992) with $T=0.3$ keV, $z=0$, a metal abundance
unconstrained but consistent with solar, and the absorption column
consistent with the \rosat\ PSPC value derived by Elbaz et al.\ (1995) for
this cluster. The observed surface brightness of the soft excess is $2\times
10^{-15}$\ergscm$\;$arcmin$^{-1}$ in the 0.5--1 keV band, within a factor of
2 of the approximate \rosat\ PSPC estimate.

On the other hand, if the observed excess is due to the temporary increase
of the particle background, it would not be vignetted in the same way as are
X-rays from the sky. If so, a better approach would be to fit a correction
function to the spectral ratio in Fig.~\ref{bgratios}{\em b}\/ and multiply
by it the nominal background model for the cluster regions. Such a function
is shown in Fig.~\ref{bgratios}{\em b} for the energy band of interest. We
tried both methods of the soft excess correction and found the results
almost identical, thus the physical nature of this component is unimportant
for the purposes of this paper. We will use the spectral modeling method
below.

When deriving the A2163 temperature profile (\S\ref{sec:tprof}), we tried to
minimize the influence of the soft background excess by discarding the data
with $E<1.2$ keV where the excess is greater than 50\% of the nominal
background (Fig.~\ref{bgratios}{\em b}), and in the remaining energy band,
using the warm gas model derived above. The 1.2 keV energy cutoff is used
for all A2163 radial bins for uniformity. Using a still higher cutoff would
further limit the background uncertainty but also decrease the statistical
accuracy of the results.

The ACIS readout artifact, although not particularly important in the
absence of sharp brightness peaks in these clusters, was subtracted as an
additional background component. It was modeled under the assumption of no
pile-up by randomizing the CHIPY coordinate of the events, recalculating the
PI values for the new coordinate, and normalizing the resulting images or
spectra by the ratio of the readout exposure (41 ms) to the useful frame
exposure (3.2 s).

\subsection{Spectral fitting}
\label{sec:spec}

To fit a spectrum in a large region, a telescope response file (ARF) was
calculated by weighting the mirror effective area within the region with the
observed cluster brightness distribution in the 0.5--2 keV band, taken as a
representation of the cluster projected emission measure (the changes in
vignetting within each region are negligible in this energy band). The
spatial nonuniformity of the CCD quantum efficiency was also included in the
ARF. For A665, observed at the focal plane temperature of $-110^\circ$C when
the CTI-induced nonuniformities were large, we used the formula by Vikhlinin
(2000). For A2163, observed at $-120^\circ$C which improved the uniformity,
we used the standard QEU maps from the calibration database. The ARFs also
included the position-independent, time-independent fudge factor of 0.93 at
$E<1.8$~keV to account for the flux discrepancy between the backside- and
frontside-illuminated chips that characterizes the current (as of March
2001) combination of the spectral response matrices and quantum efficiency
curves (Vikhlinin 2000). Response matrices (RMF) for each region were
calculated by weighting the standard set of position-dependent matrices by
the observed cluster brightness distribution within the region.

To fit the A665 spectra in large regions, we used the 0.8--9 keV energy
band, while for A2163, the 1.2--10 keV band to minimize the background
uncertainty (\S\ref{sec:bg}). A narrow interval around the mirror Ir edge
(1.8--2.2 keV) was excluded due to the calibration inaccuracies causing
significant residuals. Spectra were fit with the MEKAL model, fixing the
absorption column to the Galactic value $N_H=4.24\times 10^{20}$\cmsq\ for
A665 (Dickey \& Lockman 1990) and to the \rosat\ PSPC value $N_H=1.65\times
10^{21}$\cmsq\ for A2163 (Elbaz et al.\ 1995; this is slightly higher than
the Galactic H~{\small I} column $1.2\times 10^{21}$\cmsq). Freeing $N_H$
resulted in best-fit values consistent with the above assumptions. The
background was modeled as described in \S\ref{sec:bg}. Changes of the
best-fit temperature resulting from the $\pm5$\% change of the background
normalization were added in quadrature to the statistical uncertainties of
the temperatures to represent the 90\% background uncertainty; it has
noticeable effect only in the outer cluster regions.

\subsection{Average temperatures}
\label{sec:avg}

To check the consistency with earlier measurements, we first analyzed the
spectra in $r=6'$ regions that include most of the cluster emission. For
A665, we obtain $T=8.8\pm 0.9$ keV, in agreement with the wide-aperture
\ginga\ value ($8.2\pm 0.9$ keV, Birkinshaw, Hughes, \& Arnaud 1991) and the
\asca\ value for the same $r=6'$ region ($8.4\pm 1.5$ keV, Markevitch 1996).
For A2163, we obtain $T=12.3^{+1.3}_{-1.1}$ keV, marginally lower than the
joint \ginga\ and \rosat\ PSPC fit $14.6^{+0.9}_{-0.8}$ keV (Elbaz et al.\
1995) and in agreement with the \asca\ value for the same region ($11.5\pm
1.5$ keV, Markevitch 1996). We note here that without applying the
correction to the ARFs at $E<1.8$ keV mentioned above, the best-fit ACIS
temperatures (10.5 keV for A665 and 15.2 keV for A2163) would be
inconsistent with earlier results and the best-fit value of $N_H$ for A2163
would be unreasonably high.

%%%%%%%%%%%%%%%%%%%%%%%%%%%%%%%%%%%%%%%%%%%%%%%%%%%%%%%%%%%%%%%%%%%%%%%%%%
\begin{figure*}[tb]
\pspicture(0,10.7)(18.5,20)
%\psgrid(0,0)(18.5,20)

\rput[tl]{0}(0.5,20){\epsfxsize=8.0cm \epsfclipon
\epsffile[22 152 508 602]{665.dss.acis.ps_bw_dist}}

\rput[tl]{0}(9.5,20){\epsfxsize=8.5cm \epsfclipon
\epsffile[22 152 508 602]{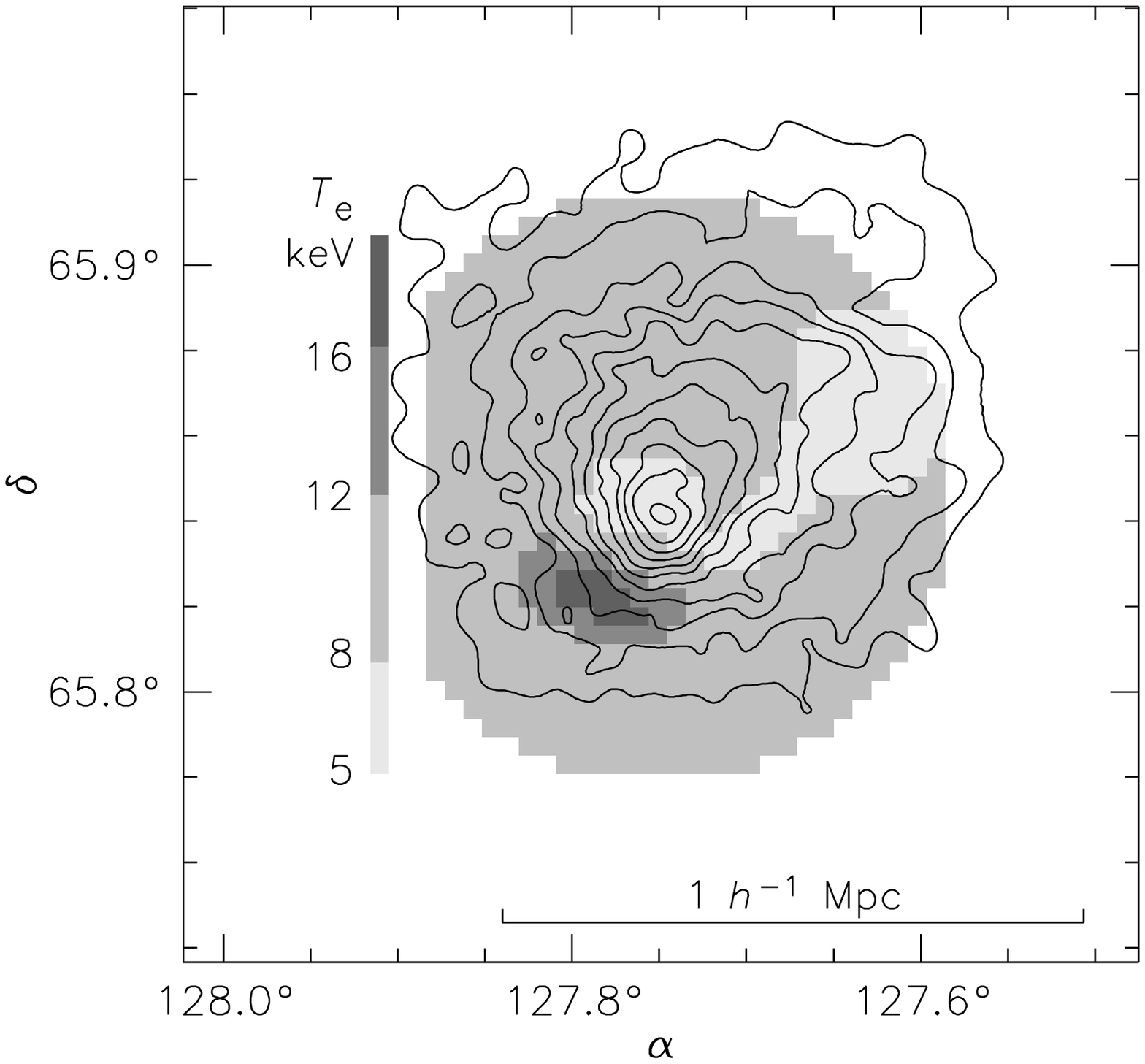}}

\rput[bl]{0}(8.1,19.2){\large\bi a}
\rput[bl]{0}(17.4,19.4){\large\bi b}

\rput[tl]{0}(-0.1,11.8){
\begin{minipage}{18.5cm}
\small\parindent=3.5mm
{\sc Fig.}~\ref{665map}.---({\em a}) A665 X-ray brightness contours (ACIS,
0.5--4 keV band) overlaid on the DSS plate. X-ray point sources were
excluded before smoothing the image. The size of the DSS plate is 8.5\am\ or
$1.0\,h^{-1}$ Mpc. ({\em b}) ACIS temperature map (grayscale) overlaid on
the X-ray contours. Different shades of gray approximately correspond to
significantly different temperatures; contours are log-spaced by a factor
$2^{1/2}$.
\par
\end{minipage}
}
\endpspicture
\refstepcounter{figure}
\label{665map}
\end{figure*}
%%%%%%%%%%%%%%%%%%%%%%%%%%%%%%%%%%%%%%%%%%%%%%%%%%%%%%%%%%%%%%%%%%%%%%%%%%

\subsection{Temperature profiles}
\label{sec:tprof}

To derive the radial temperature profiles, we extracted the spectra in
annuli centered on the cluster brightness peak (some of the annuli are
incomplete due to the limited field of view). Point sources were excluded
from the spectra. The fitting results are shown in Fig.~\ref{tprofs}. In
general, the uncertainties are large due to the short exposures.  For the
outer region in A2163, we also show the temperature obtained by ignoring the
soft background excess (black dashed line). The difference is within the
uncertainty; for the inner annuli, it is unnoticeable.

For comparison, Figure~\ref{tprofs} also includes earlier projected
temperature profiles from \asca\ (Markevitch 1996) that showed a temperature
decline in both clusters. The Markevitch (1996) analysis was limited to
$E>2.5$ keV and therefore insensitive to the anomalous soft background in
the A2163 field. An alternative \asca\ derivation of the A2163 profile by
White (2000) is also reproduced in Fig.~\ref{tprofs}{\em b}\/ (only the
interesting outer $r=4'-14'$ radial bin is shown for clarity).%
\footnote{Their $1\sigma$ interval is converted to 90\% and the cross is
centered at the emission-weighted radius rather than the middle of the bin
(at these radii, the cluster surface brightness declines with radius as
$r^{-4}$, see, e.g., Vikhlinin et al.\ 1999). As seen from
Fig.~\ref{tprofs}{\em b}, contrary to the impression that the White (2000)
paper tried to convey, the two \asca\ results did not disagree.}
For A665, that work did not present measurements at the interesting
off-center distances. For A2163, the last radial bin of the \bepposax\
profile from Irwin \& Bregman (2000) is also shown. The new ACIS
temperatures for A2163 are in agreement with the earlier values at all
radii, except the Irwin \& Bregman measurement that deviates by $\sim
2.4\sigma$.  The new A665 profile does not cover the radii where the
Markevitch (1996) \asca\ fits showed the temperature drop. The new profile
for A2163 does indicate a temperature decline at large radii, but the
uncertainty is large.  The forthcoming longer observation of A2163 with
\chandra\ and the analysis of the \xmm\ data (for preliminary results, see
Pratt et al.\ 2001) will strengthen the constraints on the radial
temperature decline in this interesting cluster.

%%%%%%%%%%%%%%%%%%%%%%%%%%%%%%%%%%%%%%%%%%%%%%%%%%%%%%%%%%%%%%%%%%%%%%%%%%
\begin{figure*}[tb]
\pspicture(0,10.3)(18.5,20)
%\psgrid(0,0)(18.5,20)

\rput[tl]{0}(0.5,20){\epsfxsize=8.0cm \epsfclipon
\epsffile[22 152 508 602]{2163.dss.acis.ps_bw_dist}}

\rput[tl]{0}(9.5,20){\epsfxsize=8.5cm \epsfclipon
\epsffile[22 152 508 602]{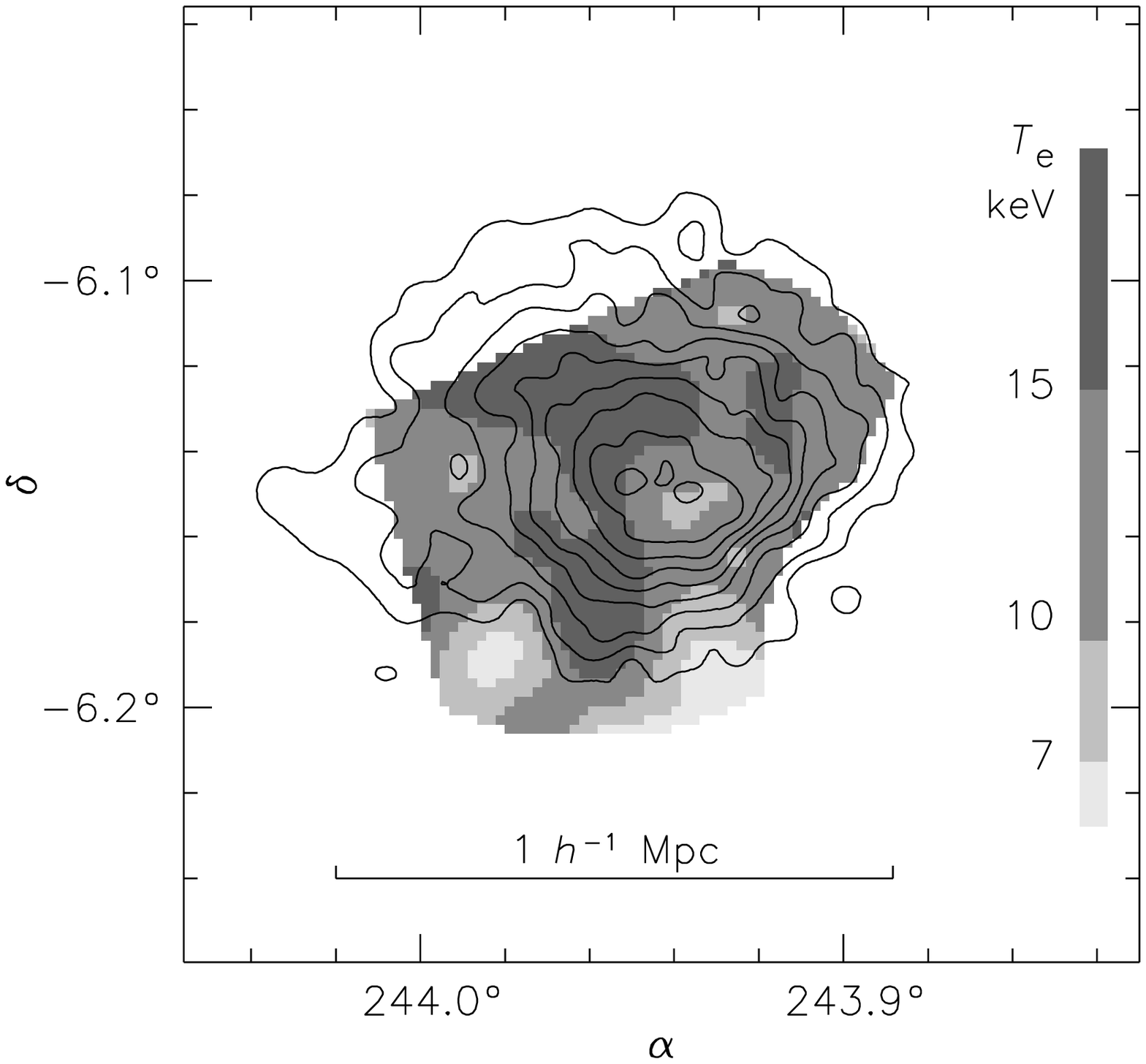}}

\rput[bl]{0}(1.5,19.3){\large\bi a}
\rput[bl]{0}(11.3,19.3){\large\bi b}

\rput[tl]{0}(-0.1,11.8){
\begin{minipage}{18.5cm}
\small\parindent=3.5mm
{\sc Fig.}~\ref{2163map}.---({\em a}) A2163 X-ray brightness contours (ACIS,
0.8--4.5 keV band) overlaid on the DSS plate.  The size of the DSS plate is
8.5\am\ or $1.1\,h^{-1}$ Mpc. X-ray point sources are excluded. ({\em b})
ACIS temperature map (grayscale) overlaid on the X-ray contours. Regions of
the map with large relative uncertainties are cut off for clarity. Different
shades of gray approximately correspond to significantly different
temperatures.  Contours are log-spaced by a factor $2^{1/2}$.
\par
\end{minipage}
}
\endpspicture
\refstepcounter{figure}
\label{2163map}
\end{figure*}
%%%%%%%%%%%%%%%%%%%%%%%%%%%%%%%%%%%%%%%%%%%%%%%%%%%%%%%%%%%%%%%%%%%%%%%%%%

\subsection{Derivation of the temperature maps}
\label{sec:tmap}

Detailed temperature maps of the cluster central regions were derived by the
technique described in Markevitch et al.\ (2000). For A665, we extracted
images in the 0.7--2.5--5.0--9.0 keV bands, while for A2163, in the
0.8--2.5--4.5--8.0 keV bands (the number of energy intervals is limited by
low statistics). Point sources were excluded.  Background maps that included
the genuine and the readout components (as described in \S\ref{sec:bg}) were
subtracted from each image. The temperature maps are limited to the
brightest regions of the clusters where the background accuracy is not as
important as it is at large radii, thus to improve statistics, we chose to
use a wider band for A2163 than that used for the profiles in
\S\ref{sec:spec}. The images were then divided by the exposure maps that
included vignetting and quantum efficiency nonuniformity (see
\S\ref{sec:spec}). The resulting images were smoothed by a Gaussian filter with
variable width, same for all energy bands. A temperature in each pixel of
the map was fitted using the data values from the smoothed images properly
weighted with their errors, fixing $N_H$ to the values given in
\S\ref{sec:spec}. The resulting temperature maps, the first for A665 and
A2163 with interesting spatial resolution, are discussed below.

\section{DISCUSSION}

Optical data for both clusters show significant structure, which may
indicate either the genuine unrelaxed state or complex projection effects.
In A665, Geller \& Beers (1982) and Beers \& Tonry (1986) noticed an
elongation in galaxy number density distribution; on the other hand,
G\'omez, Hughes, \& Birkinshaw (2000) detect ``only subtle evidence for
substructure'' in the radial galactic velocities, suggesting that any merger
activity should occur mostly in the plane of the sky. For A2163, the
velocity data reveal significant subclustering in the redshift-coordinate
space (G. Soucail, M. Arnaud, \& G. Mathez, in preparation). The X-ray data
confirm that both clusters are mergers.

\subsection{Merger in A665}
\label{sec:merg665}

The X-ray contours of A665 (Fig.~\ref{665map}{\em a}) are elongated in the
same direction as the galaxy distribution (Geller \& Beers 1982), and their
appearance alone suggests that the bright core associated with the main
galaxy concentration is flying in the southern direction with respect to the
more diffuse cluster component (this was suggested, e.g., by Jones \&
Saunders 1996 already from the \rosat\ PSPC image). Indeed, our ACIS
temperature map, shown in Fig.~\ref{665map}{\em b}, reveals a remarkable
shock in front of this flying core. The core itself is cooler than the
cluster average. A less-apparent brightness elongation towards northwest is
also cool; it likely corresponds to a remnant, and possibly a trail, of a
smaller subcluster flying in the NW direction.

Examination of the ACIS images clearly shows that the excess hard emission
at the position of the hot spot is extended. We have extracted a spectrum
from the $0.7'\times 1.5'$ elliptical region centered on the hot spot. The
fit gives a 90\% lower limit of 15 keV; the value exceeds the average
cluster temperature at the 98\% confidence. Unfortunately, the number of
photons from the hot region is insufficient to exclude a power-law component
(the spectrum is fit equally well with a power law model with a photon index
$1.0\pm0.4$).  The possibility of a power-law inverse Compton contribution
is discussed below (\S\ref{sec:ic}); here we can say that it cannot mimic
the observed high temperature. If the hot region is a shock, it should be
seen in the X-ray image as a gas density jump. The ACIS image does not show
any sharp brightness edges either at the presumed bow shock or at the
boundary of the cool core (such as those seen, e.g., in A3667 by Vikhlinin
et al.\ 2001 or in 1E0657--56 by M. Markevitch et al., in preparation),
although the image is not inconsistent with a more gradual density increase
at the expected location. This can be naturally explained, for example, by
projection effects due to a nonzero angle of the core velocity with respect
to the plane of the sky.

To summarize, A665 appears to be at a stage when two cool subcluster
remnants of very different sizes are flying apart after passing through each
other (probably with a nonzero impact parameter, to explain the survival of
both remnants). To some extent, this is similar to another merging cluster,
A2142 (Markevitch et al.\ 2000), except that we do not see the sharp
subcluster edges in A665 and do see a shock. 
% in the temperature map.

\subsection{Merger in A2163}

A \chandra\ image of A2163 is overlaid on the DSS plate in
Fig.~\ref{2163map}{\em a}. An obvious complex morphology of the image
indicates that the cluster central region is in the state of violent motion.
Neither of the three brightness peaks coincides with the maximum of the
galactic density, and one can discern apparent streams of gas and asymmetric
density gradients. As one can expect from the image, the temperature map is
also complex (Fig.~\ref{2163map}{\em b}). The temperature varies by at least
a factor of 2; there are two apparent shock regions coinciding with the gas
density enhancements, while the central density peak is cool. To assess the
statistical significance of these temperature deviations, we extracted
spectra from three interesting regions, a $200''\times 90''$ ellipse that
includes the northern hot spot, an $r=60''$ circle at the position of the
southern hot spot, and an $r=30''$ circle at the cool brightness peak.
These spectra give $T>17$ keV, $T>16$ keV, and $T=8.8^{+2.9}_{-1.8}$ keV
(90\% confidence), respectively; the first and the second hot spots deviate
upwards from the cluster average by $3.4\sigma$ and $2.8\sigma$,
respectively, and the temperature at the center is below the average by
$2\sigma$, thus the gradients are significant. The northern hot spot was
present, with a marginal significance, in the \asca\ hardness ratio map
(Markevitch et al.\ 1994); other features from that work are outside the
area covered by the ACIS map. Preliminary \xmm\ results (Bourdin 2001)
appear to show similar temperature variations.

The structure in the A2163 temperature map is too complicated to try
guessing a merger scenario (perhaps hydrodynamic simulations of the X-ray
and optical data, as in, e.g., Roettiger et al.\ 1998 and 1999, would help).
It appears that the density peak is a remnant of a cool subcluster that has
survived a merger and is now surrounded by shock-heated gas.

%%%%%%%%%%%%%%%%%%%%%%%%%%%%%%%%%%%%%%%%%%%%%%%%%%%%%%%%%%%%%%%%%%%%%%%%%%
\begin{figure*}[tb]
\pspicture(0,1.8)(18.5,11.6)
%\psgrid(0,0)(18.5,20)

\rput[tl]{0}(0.0,11.5){\epsfxsize=8.5cm \epsfclipon
\epsffile[22 152 508 602]{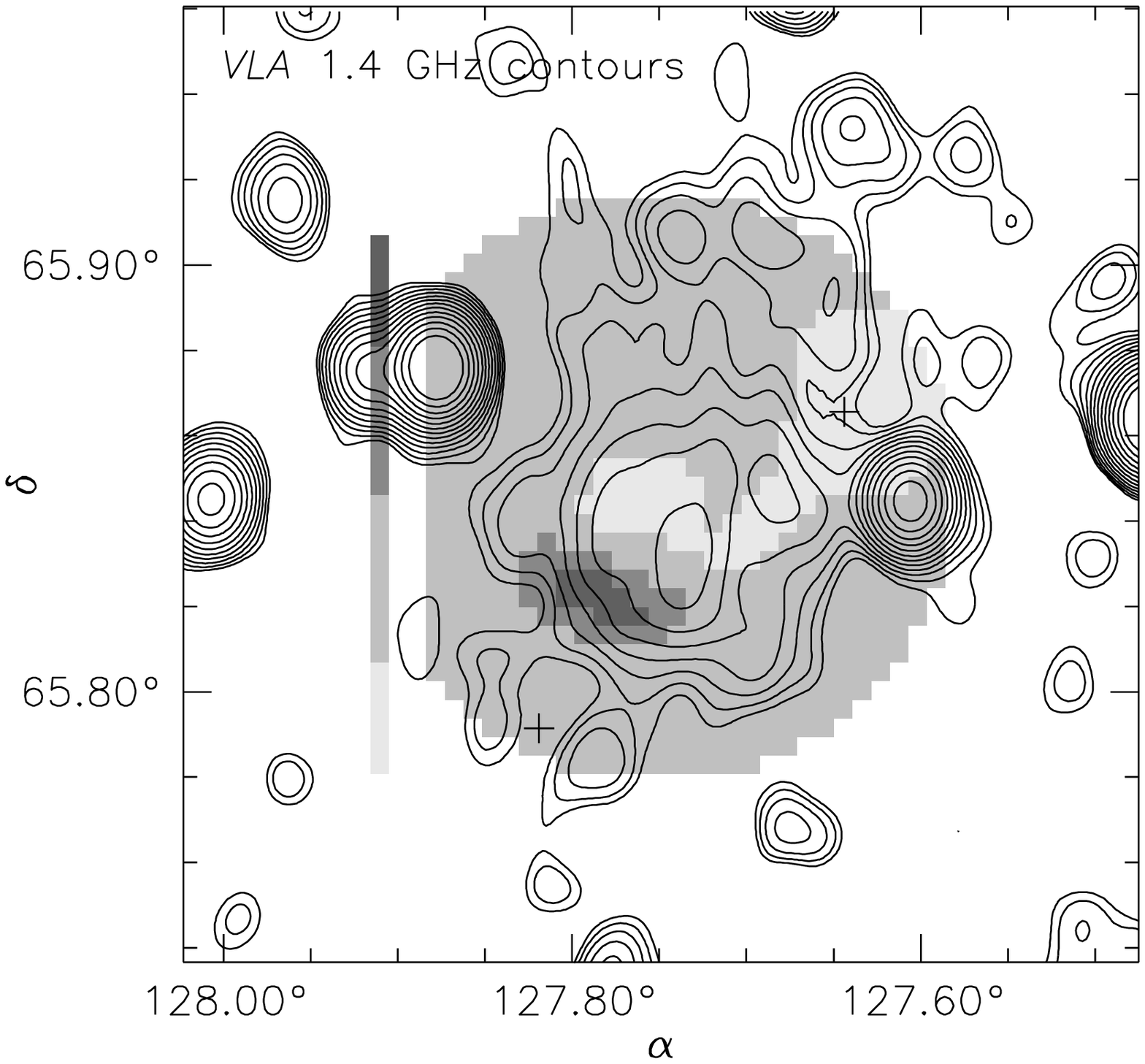}}

\rput[tl]{0}(9.5,11.5){\epsfxsize=8.5cm \epsfclipon
\epsffile[22 152 508 602]{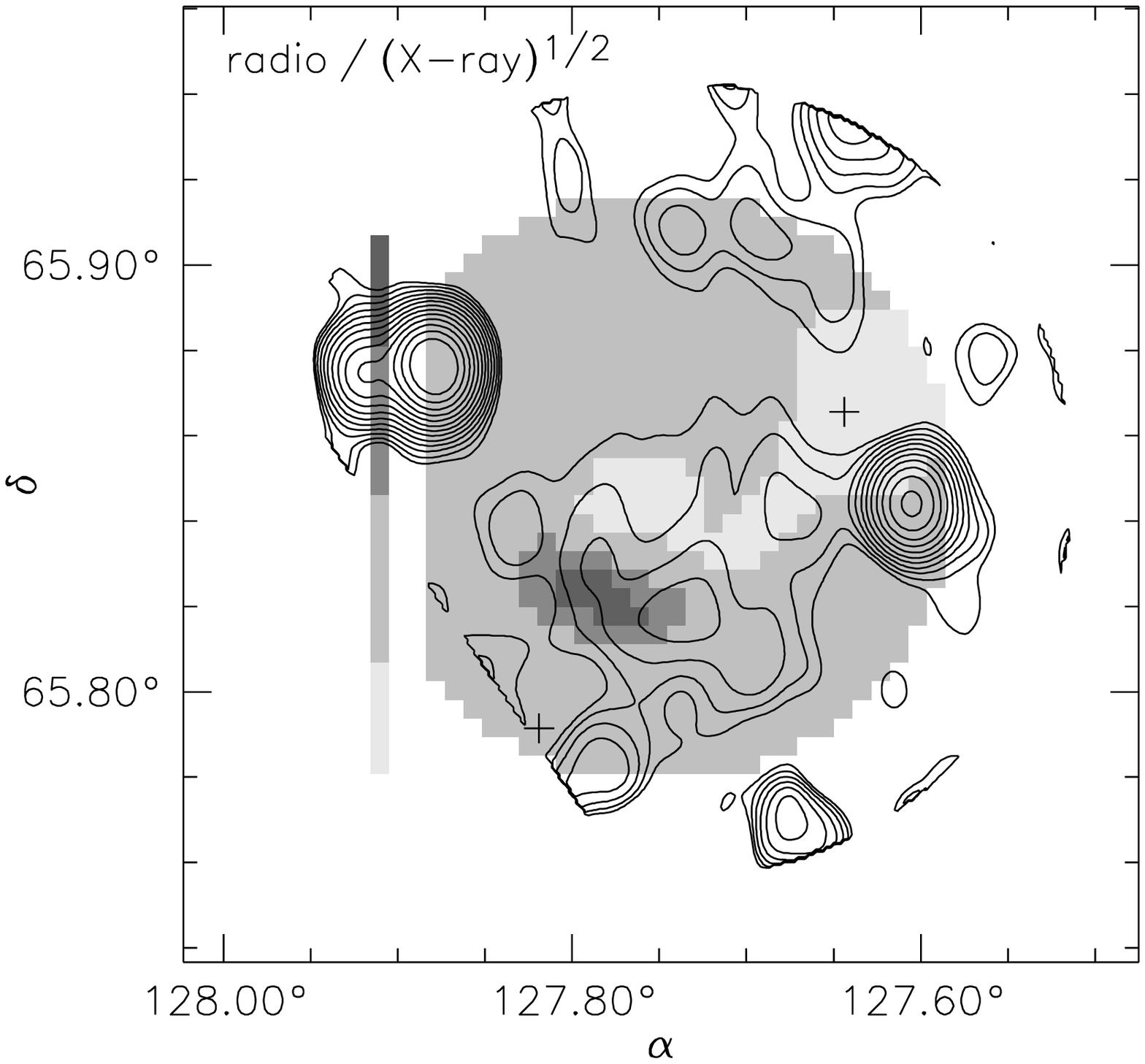}}

\rput[bl]{0}(7.9,10.9){\large\bi a}
\rput[bl]{0}(17.4,10.9){\large\bi b}

\rput[tl]{0}(-0.1,3.1){
\begin{minipage}{18.5cm}
\small\parindent=3.5mm
{\sc Fig.}~\ref{665radio}.---({\em a}) A665 temperature map from
Fig.~\ref{665map} (grayscale) overlaid on the {\em VLA}\/ 1.4 GHz contours
of the radio halo from Giovannini \& Feretti (2000).  ({\em b}) Same as
panel {\em a}\/ but contours show radio brightness divided by square root of
the X-ray brightness (edges are cut off for clarity). The two crosses
indicate positions of the removed radio point sources; the angular
resolution of the radio images is 45\as. Contours are log-spaced by a factor
of $2^{1/2}$.
\par
\end{minipage}
}
\endpspicture
\refstepcounter{figure}
\label{665radio}
\end{figure*}
%%%%%%%%%%%%%%%%%%%%%%%%%%%%%%%%%%%%%%%%%%%%%%%%%%%%%%%%%%%%%%%%%%%%%%%%%%

\subsection{Comparison with radio halos}
\label{sec:halos}

Both A665 and A2163 have luminous radio halos (A665: Moffet \& Birkinshaw
1989; Jones \& Saunders 1996; Giovannini \& Feretti 2000; A2163: Herbig \&
Birkinshaw 1994; Giovannini et al.\ 1999; Feretti et al.\ 2001; M.
Birkinshaw \& T. Herbig, in preparation). Having derived the gas temperature
maps that reveal merger shocks, we can compare these maps to the radio halo
images to determine whether the shocks indeed are the sites of the electron
acceleration as proposed by Harris et al.\ (1980) and Tribble (1993).

Figures \ref{665radio}{\em a}\/ and \ref{2163radio} show the 1.4 GHz radio
halo images overlaid on the gas temperature maps. For A665, the radio {\em
VLA}\/ image is reproduced from Giovannini \& Feretti (2000). For A2163, an
image of the central bright region of the halo is from M. Birkinshaw \& T.
Herbig (in preparation). Another {\em VLA}\/ image of A2163 has just been
presented in Feretti et al.\ (2001) from which we took the positions of the
point sources marked in Fig.~\ref{2163radio}.

In A665 (Fig.~\ref{665radio}{\em a}), our X-ray shock region coincides
with the distinct southeastern edge of the ``limb-brightened'' (Jones \&
Saunders 1996) radio halo, suggesting that this shock indeed plays a role in
the generation of the halo. The overall geometry suggests that the large
cool subcluster has arrived to the present location approximately from the
northwest, with its shock blazing a trail of relativistic electrons along
the way. These electrons are unlikely to penetrate the subcluster because of
the disjoint magnetic field structure of the moving subcluster and the
surrounding shock-heated gas, so in three dimensions, the trail would be
hollow and produce the observed limb-brightened radio halo. (Some confusion
is likely to be added by the subcluster moving at an angle to the sky plane
as suggested in \S\ref{sec:merg665}, so that most of the shock region would
be projected onto the bright cool core where the X-ray-derived temperature
is dominated by the core.) This qualitative scenario is not inconsistent
with the Jones \& Saunders' estimate of an average age of the A665 halo,
$t\lax 10^8$ yr. From the observed temperature jump across the shock (from
$T_0\approx 8$ keV to $T_1>15$ keV), the Mach number of the large subcluster
is $M>1.8$. A bow shock moving at 1.8--2.5 times the sound speed would
travel of order $150-200\,h^{-1}$ kpc during $10^8$ yr, which is comparable
to the size of the halo's brightest region. An obvious observable prediction
of this scenario would be that the radio spectrum of the halo at the present
shock position would correspond to the youngest particles, while the
``tail'' of the halo would have a spectral signature of the old particles.

\noindent
%%%%%%%%%%%%%%%%%%%%%%%%%%%%%%%%%%%%%%%%%%%%%%%%%%%%%%%%%%%%%%%%%%%%%%%%%%
%\begin{figure*}[tb]
\pspicture(0,1.2)(9,12)
%\psgrid(0,0)(9,12)

\rput[tl]{0}(0.0,11.5){\epsfxsize=8.5cm \epsfclipon
\epsffile[22 152 508 602]{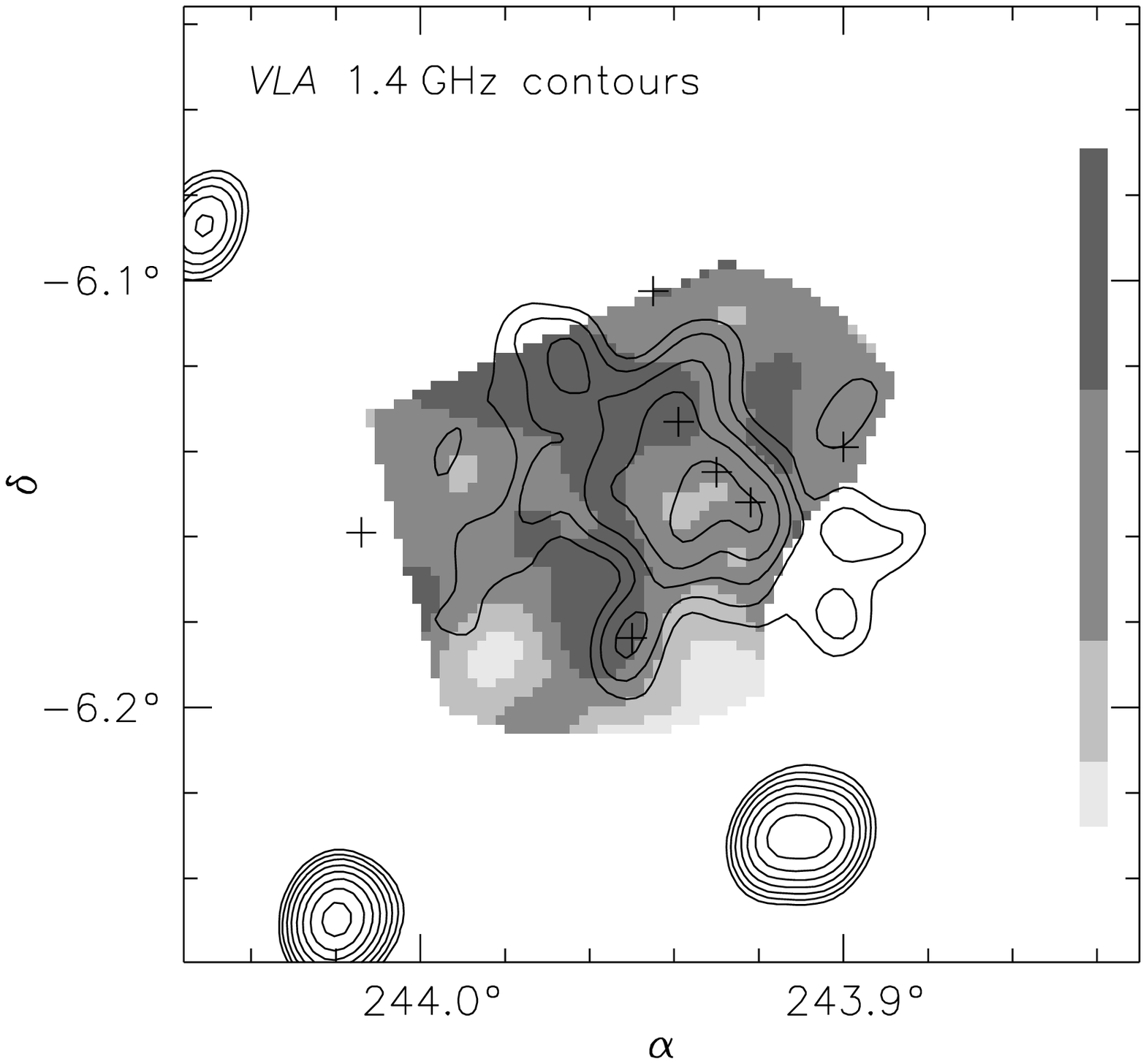}}

\rput[tl]{0}(-0.1,3.3){
\begin{minipage}{8.75cm}
\small\parindent=3.5mm
{\sc Fig.}~\ref{2163radio}.---A2163 temperature map from Fig.~\ref{2163map}
(grayscale) is overlaid on the {\em VLA}\/ 1.4 GHz contours showing the
central region of the radio halo (from M. Birkinshaw \& T. Herbig, in
preparation).  Contours are log-spaced by a factor of $2^{1/2}$. The angular
resolution of the image is 45\as. Crosses (from Feretti et al.\ 2001) show
positions of unrelated radio point sources in the central region.
\par
\end{minipage}
}
\endpspicture
\refstepcounter{figure}
\label{2163radio}
%\end{figure*}
%%%%%%%%%%%%%%%%%%%%%%%%%%%%%%%%%%%%%%%%%%%%%%%%%%%%%%%%%%%%%%%%%%%%%%%%%%
%
%
~~~In A2163 (Fig.~\ref{2163radio}), there is also an apparent correlation
between the gas temperature and the radio brightness --- the main elongation
of the halo's brightest part coincides with the hottest region of the
cluster (the southeastern radio elongation is probably due to a point
source). However, the hottest region itself coincides with an enhancement of
the X-ray brightness (Fig.~\ref{2163map}{\em b}). Feretti et al.\ (2001)
report that on larger scales and at lower radio brightness levels, the radio
emission largely follows the X-ray emission. So for A2163, there is an
ambiguity of whether the X-ray brightness or the gas temperature underlies
the radio morphology. A similar observation of the large-scale similarity of
the X-ray and radio brightness was made for several other clusters with
well-studied halos (e.g., Deiss et al.\ 1997; Liang et al.\ 2000; Govoni et
al.\ 2001).

However, for A665 with a more transparent merger geometry, it is clear that
it is not simply the X-ray brightness, or the projected gas density, that
determines the synchrotron brightness. Figure~\ref{665radio}{\em b}\/ shows
the radio image divided by the square root of the X-ray brightness (to do
this, we smoothed the X-ray image to the resolution of the radio image). A
square root of an X-ray image crudely represents the projected distribution
of the number density of thermal electrons. Therefore, such a ratio has an
approximate physical meaning of the synchrotron brightness (or the number of
relativistic electrons if the magnetic field is reasonably uniform) per
thermal electron on the line of sight. As seen in Fig.~\ref{665radio}{\em
b}, this quantity peaks at the position of the shock. If the seed particles
for acceleration are distributed more or less uniformly in the cluster
thermal gas (as they are expected to be, for example, if the seeds come from
the ICM pool as proposed by Liang et al.\ 2000), then the image of this
ratio in A665 indicates that the particle acceleration indeed occurs at the
shock site.

Several sources of the moderately high-energy or relativistic seeds for
acceleration to the ultra-relativistic energies were proposed (see, e.g.,
En{\ss}lin 2000 for a review). Besides the radio galaxies (Jaffe 1977) and
the high-energy tail of the ICM electrons (Liang et al.\ 2000), they may
include particles heated by mild turbulence induced by galaxy motions (e.g.,
Deiss et al.\ 1997) or past mergers (e.g., Sarazin 2000). A discussion of
these sources and the precise mechanism of acceleration is beyond the scope
of this paper. Our results appear to indicate that in A665 and possibly in
A2163, the final acceleration of the seed electrons to ultra-relativistic
energies occurs inside the shock-heated gas. 

It is worth pointing out here that a merging cluster has to be in a
favorable projection for a clear view of the shocks (as in A665), so one
expects that in many clusters seen at less favorable angles, the temperature
features and their association with the radio brightness would be less
apparent.

\subsection{Estimate of inverse Compton flux}
\label{sec:ic}

The population of the relativistic electrons producing synchrotron radio
halos should also scatter cosmic microwave background photons to X-ray
energies via the inverse Compton (IC) mechanism. A tentative detection of
the hard IC emission was reported for some clusters (Fusco-Femiano et al.\
1999, 2000). One may ask whether the above correlation between the
temperature maps (which are, essentially, the X-ray hardness ratio maps) and
the radio emission could be trivially explained by an IC contribution. The
spectral index of the X-ray IC emission should be the same as the radio
synchrotron index. Radio spectra of halos are steep; from the Jones \&
Saunders (1996) measurements at 1.4~GHz and 151~MHz, the energy index of the
A665 halo is $\alpha_r \simeq 0.8$ (this is slightly less steep than a
typical slope; they also present a higher frequency point where the spectrum
steepens, but the lower frequencies are more relevant for the IC estimate).
The expected IC spectrum ($\alpha_x = \alpha_r \simeq 0.8$) is much softer
than the additional spectral component needed in our energy band to mimic
the hot spot in A665 ($\alpha_x < 0$, which would require an unphysical
power spectrum of the relativistic electrons). Thus the hot spot in A665
and, similarly, the hot regions in A2163, cannot be due to the IC
contribution.

It is interesting to try and place a more accurate limit on the possible IC
flux in the A665 hot spot region, since it is the place where the ratio of
the radio to X-ray emission is highest. We can estimate the expected IC flux
(as in, e.g., Sarazin 1988) using the estimates of the equipartition
magnetic field and spectral index from Jones \& Saunders (1996) and a
1.4~GHz flux from the Giovannini \& Feretti (2000) radio map.  The X-ray
spectrum cannot exclude such a power-law component at a level up to two
orders of magnitude above the expected flux. Thus, we cannot place
interesting constraints on the IC contribution.

\section{SUMMARY}

Using \chandra\ data, we have derived gas temperature maps for the A665 and
A2163 clusters. Both show strong temperature variations indicating ongoing
mergers. In A665, we discover a bow shock in front of the cluster core
indicating that the core is moving with a relatively high Mach number. In
A2163, the temperature map is too complicated to allow unambiguous
interpretation. For the first time, we compare cluster temperature maps with
radio halo images and show that the hottest regions correlate with the
brightest regions of the radio halos. This indicates that acceleration of
the relativistic electrons that generate the radio halos occurs at the sites
of shock heating.

\acknowledgements

We are grateful to Leon VanSpeybroeck for allowing us to analyze his
Guaranteed Time observations of A665 and A2163, and to Mark Birkinshaw for
providing us the A2163 radio image prior to publication. We also thank them,
Dan Harris and David Buote (the referee) for useful comments and
discussions. We are grateful to Drs.\ G. Giovannini and L. Feretti for
sending us their A665 radio image in electronic format. The results
presented here are made possible by the successful effort of the entire
\chandra\ team to build, launch and operate the observatory.  Support for
this study was provided by NASA contract NAS8-39073 and grant NAG5-9217.

%%%%%%%%%%%%%%%%%%%%%%%%%%%%%%%%%%%%%%%%%%%%%%%%%%%%%%%%%%%%%%%%%%%%%%%%%%%

%%%%%%%%%%%%%%%%%%%%%%%%%%%%%%%%%%%%%%%%%%%%%%%%%%%%%%%%%%%%%%%%%%%%%%%%%%%


\begin{references}

\reference{} Beers, T. C., \& Tonry, J. L. 1986, ApJ, 300, 557

%\reference{} Birkinshaw, M., \& Herbig, T., 2001, in preparation

\reference{} Birkinshaw, M., Hughes, J. P., \& Arnaud, K. A. 1991, ApJ, 379,
466

\reference{} Bourdin, H., Slezak, E., Bijaoui, A., \& Arnaud, M. 2001, in 
Galaxy Clusters and the High Redshift Universe Observed in X-rays, XXXVI
Recontres de Moriond, in press (astro-ph/0106138)

\reference{} Buote, D. A. 2001, ApJ, in press (astro-ph/0104211)

\reference{} Buote, D. A. \& Tsai, J. C. 1996, ApJ, 458, 27 

\reference{} Deiss, B. M., Reich, W., Lesch, H., \& Wielebinski, R. 1997,
A\&A, 321, 55

\reference{} Dolag, K., \& En{\ss}lin, T. A. 2000, A\&A, 362, 151

\reference{} Dickey, J. M., \& Lockman, F. J. 1990, ARA\&A, 28, 215

\reference{} Elbaz, D., Arnaud, M., \& B\"ohringer, H. 1995, A\&A, 293, 337

\reference{} En{\ss}lin, T. A. 2000, in The Universe at Low Radio
Frequencies, ASP Conference Series, in press (astro-ph/0001433)

\reference{} Feretti, L. 2000, in The Universe at Low Radio
Frequencies, ASP Conference Series, in press (astro-ph/0006379)

\reference{} Feretti, L., Fusco-Femiano, R., Giovannini, G., \& Govoni, F.
2001, A\&A, in press (astro-ph/0104451)

\reference{} Fusco-Femiano, R., dal Fiume, D., Feretti, L., Giovannini, G.,
Grandi, P., Matt, G., Molendi, S., \& Santangelo, A. 1999, ApJ, 513, L21

\reference{} Fusco-Femiano, R., et al.\ 2000, ApJ, 534, L7 

\reference{} Geller, M. J., \& Beers, T. C. 1982, PASP, 94, 421

\reference{} G{\'o}mez, P. L., Hughes, J. P., \& Birkinshaw, M. 2000, ApJ,
540, 726

\reference{} Giovannini, G., \& Feretti, L. 2000, New Ast., 5, 335

\reference{} Govoni, F., En{\ss}lin, T. A., Feretti, L., \& Giovannini, G.
2001, A\&A in press (astro-ph/0101418)

\reference{} Giovannini, G., Tordi, M., \& Feretti, L. 1999, New Ast., 4, 141

\reference{} Harris, D. E., Kapahi, V. K., \& Ekers, R. D. 1980, A\&AS, 39,
215 

\reference{} Henry, J. P., \& Briel, U. G. 1995, ApJ, 443, L9

\reference{} Herbig, T., \& Birkinshaw, M. 1994, American Astronomical
Society Meeting, 185, 5307

\reference{} Irwin, J. A., \& Bregman, J. N. 2000, ApJ, 538, 543 

\reference{} Jaffe, W. J. 1977, ApJ, 212, 1

\reference{} Jones, M., \& Saunders, R. 1996, in R\"ontgenstrahlung from
the Universe, ed.\ H. U. Zimmermann et al.\ (MPE Report 263), 553

\reference{} Kaastra, J. S. 1992, ``An X-Ray Spectral Code for Optically Thin
Plasmas'' (Internal SRON-Leiden Report, updated version 2.0)

\reference{} Liang, H., Hunstead, R. W., Birkinshaw, M., \& Andreani, P.
2000, ApJ, 544, 686

\reference{} Markevitch, M.\ 1996, ApJ, 465, L1

\reference{} Markevitch, M. 2001, \chandra\ calibration memo,\\
http://asc.harvard.edu/cal/, ``ACIS,'' ``ACIS Background''

\reference{} Markevitch, M., et al.\ 2000, ApJ, 541, 542 

\reference{} Markevitch, M., Sarazin, C. L., \& Vikhlinin, A. 1999, ApJ,
521, 526

\reference{} Markevitch, M., Yamashita, K., Furuzawa, A., \& Tawara, Y.\
1994, ApJ, 436, L71

\reference{} Moffet, A. T., \& Birkinshaw, M. 1989, AJ, 98, 1148 

\reference{} Neumann, D. M., et al.\ 2001, A\&A, 365, L74

\reference{} Pratt, G., Arnaud, M., \& Aghanim, N. 2001, in 
Galaxy Clusters and the High Redshift Universe Observed in X-rays, XXXVI
Recontres de Moriond, in press (astro-ph/0105431)

\reference{} Roettiger, K., Burns, J. O., \& Stone, J. M. 1999, ApJ, 518, 603

\reference{} Roettiger, K., Stone, J. M., \& Mushotzky, R. F. 1998, ApJ,
493, 62 

\reference{} Sarazin, C. L. 1988, X-ray Emission from Clusters of Galaxies
(Cambridge: Cambridge University Press)

\reference{} Sarazin, C. L. 1999, ApJ, 520, 529

\reference{} Sarazin, C. L. 2000, in Constructing the Universe with Clusters
of Galaxies, ed.\ F. Durret \& D. Gerbal, in press (astro-ph/0009094)

\reference{} Tribble, P. 1993, MNRAS, 263, 31

\reference{} Vikhlinin, A. 2000, \chandra\ calibration memo,\\
http://asc.harvard.edu/cal/Links/Acis/acis/Cal\_prods/qe/12\_01\_00

\reference{} Vikhlinin, A., Forman, W., \& Jones, C. 1999, ApJ, 525, 47

\reference{} Vikhlinin, A., Markevitch, M., \& Murray, S. S. 2001, ApJ, 551,
160

\reference{} White, D. A. 2000, MNRAS, 312, 663

\end{references}
\end{document}